# A Hybrid Data-Driven Web-Based UI-UX Assessment Model

Ebenezer Agbozo[1 (0000-0002-2413-3815)]

[1]eagbozo@urfu.ru, Department of Big Data Analytics and Methods of Video Analysis, Ural Federal University, Yekaterinburg, Russia

**Abstract**

Today, a large proportion of end user information systems have their Graphical User Interfaces (GUI) built with web-based technology (JavaScript, CSS, and HTML). Some of these web-based systems include: Internet of Things (IOT), Infotainment (in vehicles), Interactive Display Screens (for digital menu boards, information kiosks, digital signage displays at bus stops or airports, bank ATMs, etc.), and web applications/services (on smart devices). As such, web-based UI must be evaluated in order to improve upon its ability to perform the technical task for which it was designed. This study develops a framework and a processes for evaluating and improving the quality of web-based user interface (UI) as well as at a stratified level. The study develops a comprehensive framework which is a conglomeration of algorithms such as the multi-criteria decision making method of analytical hierarchy process (AHP) in coefficient generation, sentiment analysis, K-means clustering algorithms and explainable AI (XAI).

1. Introduction

It has been established that more than three billion people use the Web all over the world as reported by a Lifewire article by Fischer T. (Major and Gomes 2021). As conversations regarding Web 3.0 and the metaverse deepen – especially during the last quarter of 2021, one can no longer avoid the fact that the web will play an even more central role in every facet of livelihood. Innovations such as: remote learning, remote work, online gaming, transportation, telemedicine and telehealth, food and hospitality ordering and delivery systems, fintech, blockchain-based systems, the metaverse and many more, have driven the push towards web-based technologies (Griffith 2021; Sönmez and Çakir 2021).

The rise in web-based systems as part of everyday livelihood has also raised concerns for research and innovation in security, privacy-by design, data protection, and user experience – just to name a few. All these areas have had some form of extensive research contributing to the current state of the art, but as the web-based ecosystem evolves, there is a need to fill in the gaps so as to proactively prevent future repercussions.

Presently, a huge proportion of information systems have a graphical user interface (GUI) created using web-based technologies (precisely HTML, CSS, and JavaScript). Examples of these web-based technologies include: e-commerce websites, e-learning platforms, healthcare and affective computing systems, decentralized applications (dApps), online gaming, infotainment system (in vehicles), automated teller machines (ATM), interactive display screens (for digital menu bars, information kiosks, digital signage displays at bus stops or airports, ATMs, etc.), and web apps/services (on smart devices and in internet of things (IOT)).

The World Wide Web (WWW) has come a long way since its introduction by Sir. Tim Berners-Lee in 1991 (The history of the Web - W3C Wiki n.d.). He was in favor of the rule of least power – where the less powerful a language is, the easier it is to analyze and reuse data represented in the language – which was a motivating factor for HyperText Markup Language (HTML) to be designed as it still presently is (Batalas, Khan, and Markopoulos 2021). Likewise Cascading Style Sheets (CSS), was also designed without the inclusion of conditional blocks, functions, variables, and loops, so as to ease the burden on programmers with minimal skillset (Queirós 2018). Presently, most web-based technologies are built on HTML 5 and CSS 3 due to their cross-platform advantage and the ability to achieve responsive web design (Li and Zhang 2019). Figure 1 illustrates the evolution of the web.

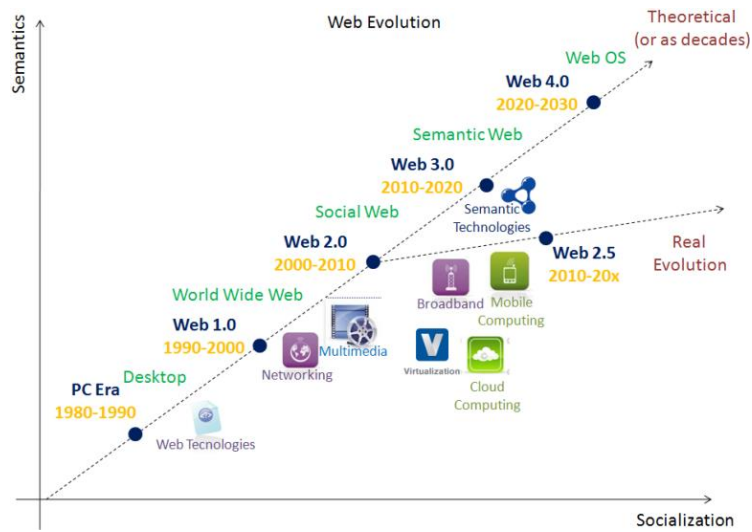

Fig. 1. The Evolution of the Web (Pileggi, Fernandez-Llatas, and Traver 2012)

In 2020, Stack Overflow reported in their Developer Survey where 57,378 software developers were surveyed, that the two (2) most commonly-used programming languages in the world are JavaScript (constituting 67.7% of the respondents) and HTML/CSS (constituting 62.4% of the sample size) (Stack Overflow Developer Survey 2020 n.d.). Thus, making web-based technology the most sought out language by developers. JavaScript's sporadic growth rate has been linked to the wider growth of the developer community, its appeal on both the server and client side, as well as professionalization (So 2018).

Every web-based information system is designed by developers (who may be users of the system) and has a user interface (web-based UI); in order for users to perform desired tasks such as communicating, or interacting with the system, they have to interact through the web-based UI to the system. Studies have highlighted that, web-based systems are information systems that are mediums for users to perform tasks they were designed for. For example web-based information systems designed to manage city transportation, provide financial literacy, managing production and customer orders, tourism services provision, pharmaceutical drug inventory management, IOT-based early fire warning systems, water monitoring systems, ICU dashboards for healthcare monitoring, and many more (Gotseva, Tomov, and Danov 2020; Iskandar and Firdaus 2018; de Morais Barroca Filho et al. 2020; Satria et al. 2018; Scott et al. 2021; Sisharini and Sunaryati Hardiani 2020; Soegoto and Pamungkas 2018; Zhukovskyy et al. 2021).

The chief aim of UI is to provide user satisfaction through the accomplishment of technical tasks for which they are designed. User experience (UX) refers to the overall experience of a user from the interaction with a technology product (Papanikou 2019). Poorly designed systems influence user effectiveness and efficiency which in turn defeats productivity envisioned for the system by its developers (Read, Tarrell, and Fruhling 2009). Thus, it can be inferred that, UX is highly dependent on user profiles (Soui et al. 2020).

The aim of this study is to develop a hybrid UI-UX quality evaluation framework. As compared to existing literature, this study focuses on incorporating all stakeholders in improving the UI-UX of any given web information system.

## 2. Web User Interface and User Experience

ISO 9241-210:2019 defines user experience (UX) as user's perceptions and responses that result from the use and/or anticipated use of a system, product or service (ISO/TC 159/SC 4 2019). As highlighted by the technical document, user perception encompasses: emotions, beliefs, preferences, perceptions, comfort, behaviors, and accomplishments that occur before, during and after use (ISO/TC 159/SC 4 2019). The report further indicates that UX is a product of a user's internal and physical state resulting from prior experiences, attitudes, skills, abilities and personality; and from the context of use

User Interface (UI) evaluation is an essential component of the software development lifecycle – as a means of providing feedback to developers and system administrators or analysts in order to improve upon the user experience. UI evaluation is essential in improving communication between users and service providers (Charfi, Ezzedine, and Kolski 2015). Studies have recommended that UI evaluation is conducted early so as to offer designers the opportunity of obtaining feedback with regards to their design ideas and aid in proceeding to interface redesign (Demetriadis, Karoulis, and Pombortsis 1999). The goal of web-based UI evaluation is to:

- estimate a web-based information system's productivity and performance;
- establish the components leading to issues or errors;
- improve and optimize the system's interactivity, efficiency and productivity;
- understand user behavior and user experience so as to leverage them for designing interactive systems (Charfi and Ezzedine 2014);

Wilson identified five (5) categories of UI evaluation techniques: Heuristic Evaluation, The Individual Expert Review, Perspective-Based UI Inspection, Cognitive Walkthrough, Pluralistic Usability Walkthrough, and Formal Usability Inspections (Wilson 2013). Another study categorized UI evaluation into three (3) groups namely: Heuristic Evaluation, Cognitive Walkthrough, and Perspective-based inspection (Hussain et al. 2018). Heuristic Evaluation, Cognitive Walkthrough, and Action Analysis were also identified as categories of UI evaluation methods (Paz et al. 2015).

In their study, researchers sought to understudy (pre-activity and post-activity) the effects of resolving Montessori activities (for example how to set the dining table) for teaching 4–5 year old preschoolers via tangible physical objects on the children's short-term retention skills and system usability (Chettaoui et al. 2022). By conducting the study, the end result aided in improving the proposed educational framework which is the aim of UX research.

With respect to previous studies on UI and UX evaluation, a number of frameworks and algorithms have been proposed and developed including: Metric-based Usability Evaluation (INUIT) - (Speicher, Both, and Gaedke 2013); UI evaluation with USE model (Fatta and Mukti 2018); Multi-objective optimization (Automated) (Soui et al. 2020); Image-Based UI Analysis with Feature-based Neural Networks (Bakaev et al. 2022); Deep features extraction for UI Evaluation (Automated) (Soui et al. 2022); Domain Ontology (Perminov and Bakaev 2019); Metric-based assessment of web user interface (WUI) quality attributes (Bakaev et al. 2018). Algorithms and methods proposed and implemented for evaluation of UI possess their value and serve various positive purposes, yet are not without limitations. Studies pointed out the following limitations:

- In measuring usability, a number of criteria are susceptible to the lack of objectivity (Bañón-Gomis, Tomás-Miquel, and Expósito-Langa 2014).

- The sole focus on user testing tends to result in usability violations that are overlooked by users with little to no HCI expertise (Cho et al. 2022).

- The cognitive walkthrough method is incapable of evaluating efficiency, attractiveness, and user satisfaction (Ambarwati and Mustikasari 2021).

- Simulation methods are incapable of replacing user-centered approaches of UI design and evaluation(Zhang et al. 2021).

- Many existing research have relied on subjective evaluation approaches of aesthetic defects that depend on feedback from end-users that makes manual evaluation techniques UI time-consuming and error-prone (Soui et al. 2020).

- Expert-based approaches are unable to detect problems effectively because they are unable to adequately capture multiple contextual factors that influence user interactions with information system within real settings (Bertini et al. 2006).

To provide solutions to limitations and shortcomings indicated such as a lack of user-centric frameworks and the sole-dependence on singular modes of evaluation, this research aims to provide a hybridization of both automated and user-based approaches to evaluate web-based UI-UX quality as well as contribute to decision making in web design and development.

### 3. Theoretical Underpinning of Proposed Model

The research proposition of this research is theoretically underpinned by the Socio-Technical Theory (STT). STT as per research purports that new technology diffuses via the interaction between social groups (primarily users and developers for this study) as well as rules (principles surrounding system interaction and user experience) (Geels 2004). STT is a theory that hinges on institutional theory, sociology and innovation studies. STT consists of interconnected constructs: technological systems; rules and institutions; and social groups, human actors, and organizations. Interaction between constructs is crucial to the diffusion, development and improvement of a given information system. Figure 2 represents the schema of the STT as proposed by Bostrom & Heinen from the perspective of web-based UI-UX (Bostrom and Heinen 1977).

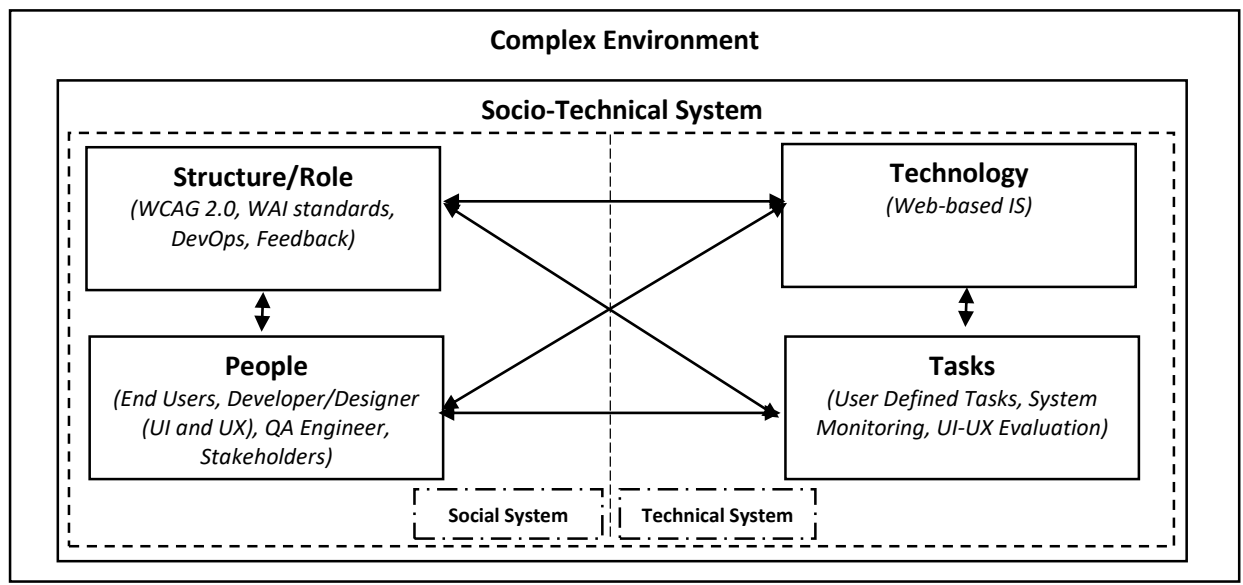

Fig. 2. Socio-Technical Theory (STT) from the perspective of Web-based UI-UX

Research defined STT's interaction and interdependency of constructs as a means of system optimization and performance maximization (Cartelli 2007; Shmelova et al. 2018). A major reason why technology failure exists can be linked to the fact that developers consider technology as a means to an end rather than a component of a socio-technical system (Kling and Lamb 1999). In light of this perspective, it is recommended that web-based UI stakeholders (owners, designers and developers) view their platforms as a socio-technical system. As such web-based UI design and redesign should aim at influencing all subsystems, and system monitoring and evaluation should be geared towards the attainment of positive UX thus ensuring all subsystems are working in harmony. As such this research proposes a hybrid UI-UX evaluation model in order to ensure balance between end users task completion efficiency and system aesthetics.

## 4. Proposed UI Evaluation Model

The study adopted a hybrid method for evaluating web user interface due to the fact that previous studies (based on literature review results) have solely focused on evaluating web user interfaces based on user experience surveys (which focuses solely on the usability component), A/B testing with eye tracking (which require experts and specialized tools or software) or user experience comparison, accessibility tests, and web performance tests. As such, the study developed a more comprehensive approach to measuring all essential components with respect to web user interface quality. This was based on three (3) key performance indicators or metrics web-usability ($U_i$) (user-based evaluation approach) - , web-accessibility ($A_i$) and web-performance ($P_i$) (both automated evaluation approaches – as represented below:

$$Web\ UI - UX\ Quality\ Metrics_i = \{P_i, A_i, U_i\} \qquad (1)$$

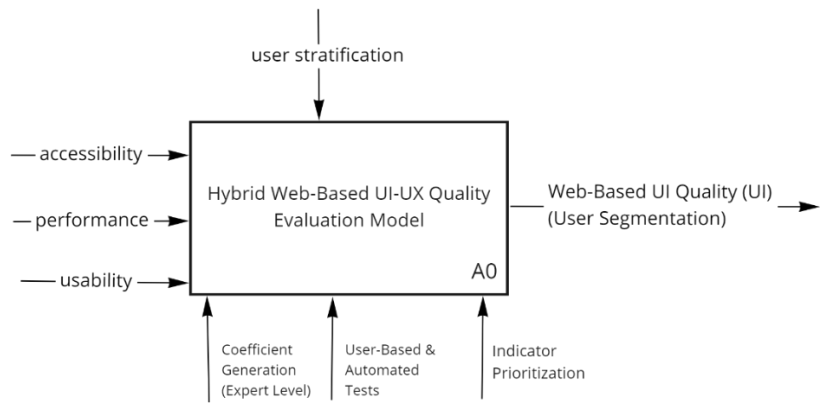

Fig. 3. Structured Analysis and Design Method (SADT) – IDEF0 Notation of the Research Model (Level A0)

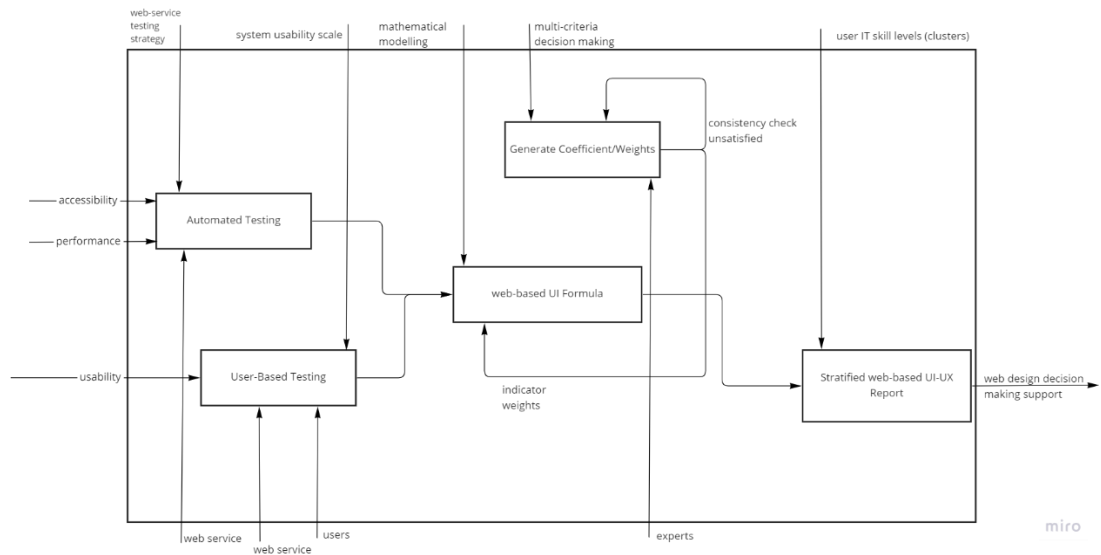

Fig. 4. Structured Analysis and Design Method (SADT) – IDEF0 Notation of the Research Model (Level A1)

The formalization of a hybrid unified framework of the model for evaluating web user interface quality is highlighted in figures 3 and 4. The structured analysis and design technique (SADT) is adopted to describe the system, process, technology and the hierarchy of functions of the proposed framework. Utilizing SADT as a systems analysis technique is aimed at establishing the framework specifications in order to inhibit algorithm failure and improve the quality. The context diagram A0 (Figure 3) is decomposed into a sub process A1 (Figure 4) which represents all components of the proposed unified framework. From A0, input is represented by the measurable metrics web accessibility, web performance, and web usability.

Figure 5 highlights the overall hybrid framework for evaluating Web UI-UX quality. It is comprised of automated analytics – expert survey (a foundation for mathematical modelling of the web UI-UX formula's coefficients); web accessibility and performance analysis; an extended usability analysis (with usability segmentation with k-means clustering algorithm and explainable AI for understanding usability segmentation results in order to improve system usability).

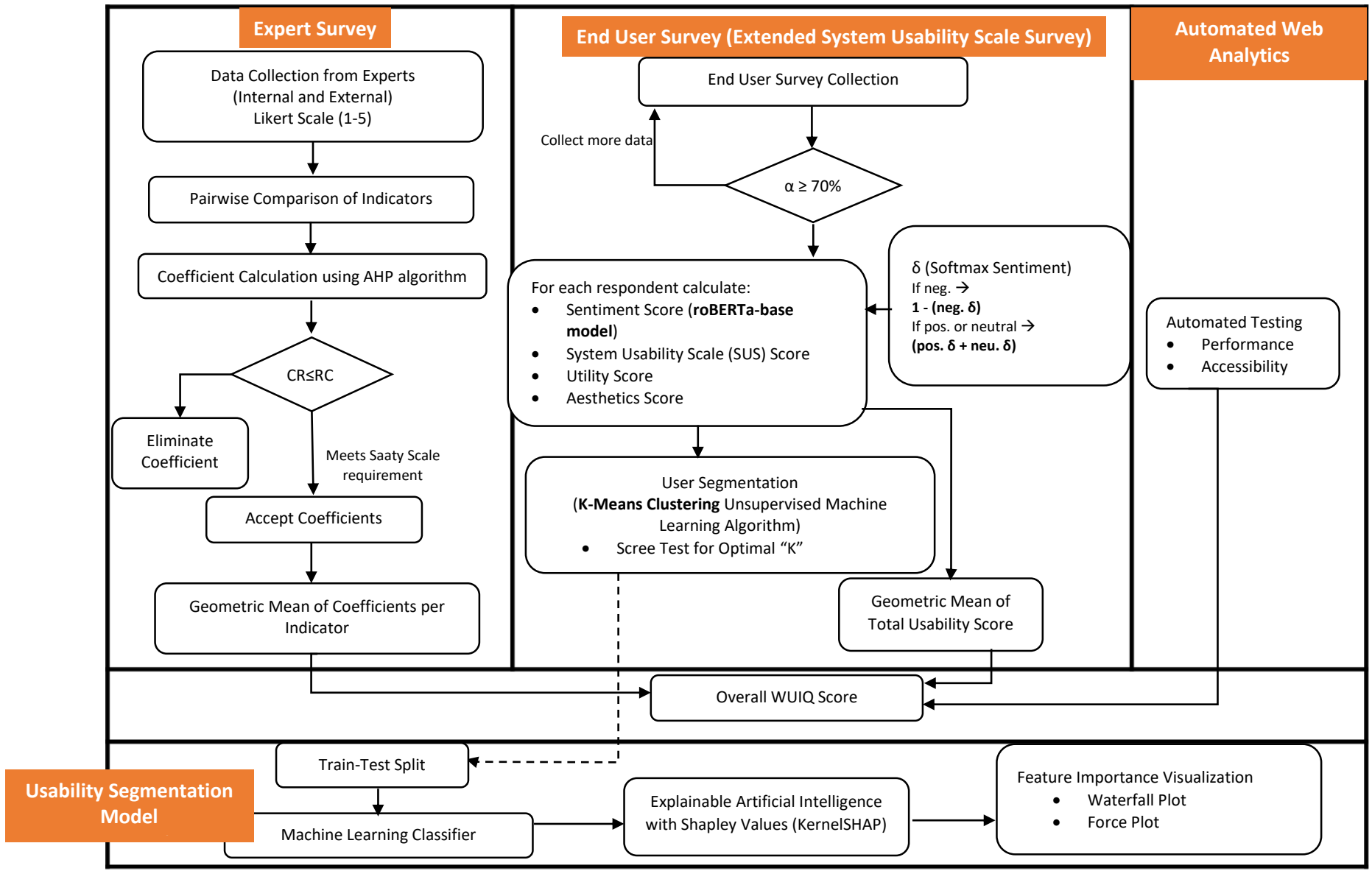

Fig. 5. Overall UI-UX Evaluation Framework

**4.1 Phase 1 – Automated Evaluation (Performance and Accessibility)**

*Web Accessibility:*

The first metric (also automatically evaluated), web accessibility which is defined as the speed at which a page loads and renders content to the user or a web page/site's ability to provide sufficient efficiency via a reasonable amount of resources. Web accessibility is defined as the ability for content to be adaptable to the needs and preferences of individual users; and also the as the ability of users to universally access web-based services and obtain necessary information (Abou-Zahra, Brewer, and Cooper 2018; Kamal et al. 2016). It is supposed to be a primary component of web development as it allows you to provide equal access to web resources (Редкокош and Косова 2020). Web accessibility usually caters for the visual, auditory, physical, speech and neurological impairments of people (Шумская 2017). It contributes to the social inclusion of people with disabilities, as well as other categories of people (Рушанян 2020). A study noted that web accessibility depends not only on the accessibility of content, but also on the accessibility of web browsers and other user components (Осипов and Графкин 2018). Another research identified certain factors that challenge accessibility, including slow internet, poor technical equipment, outdated software, low resolutions in output device, etc. (Perminov and Bakaev 2019).

Studies have analyzed the web accessibility of official websites of the universities within the Volga Federal District in Russia with primary focus on students with disabilities and visual impairments (Медведева and Ольхина 2020). Other studies have also evaluated the accessibility of e-government platforms (Cisneros, Huamán Monzón, and Paz 2021). In their study, on the seven (7) challenges of HCI (though focused on technology-augmented environments can be extended to web-based UI), Accessibility was one of the highlighted challenges. It was also highlighted as impossible and impractical to fulfil the design requirements for a web-based UI in order for it to be accessible by all people (mostly elders and persons with disabilities) regardless of their limitations (Stephanidis et al. 2019).

Accessibility solutions are not one-size-fits-all and with respect to the methods, techniques, and tools as a result of advancements in HCI, research pointed out the following existing challenges at hand which motivated this study: Development of appropriate user models; Classification and analysis of the appropriateness of various solutions for the different combinations of user and environment (contextual) characteristics; and Novel evaluation methodologies (Stephanidis et al. 2019). The sub-components of the web performance metric adopted from Web Content Accessibility Guidelines (WCAG) 2.0 include: Principle 1 - Perceivable; Principle 2 - Operable; Principle 3 - Understandable; and Principle 4 - Robust. As reported by research, a challenge lies in the quantification of web performance due to the diversity of web pages, the heterogeneity of devices and browsers, the choice of metrics (network-centric, browser-centric, as well as user-centric metrics), and the lack of effective standards. As such the goal of this research is to fill in the gap of a lack of standard testing methods and incorporate the social context in testing for user accessibility to improve the UI of web-based technologies.

*Web Performance:*

The second metric (automatically evaluated), web performance, is defined as the ability for content to be adaptable to the needs and preferences of individual users; and also the as the ability of users to universally access web-based services and obtain necessary information. Aside the structure of code (the development stack), past studies have indicated that web accessibility is dependent on factors such as internet speed, outdated software, and output device resolution. Web UI is built to perform specific tasks; for example checking out in an e-commerce web application, conducting an online examination on an e-learning platform, etc. Performance of such web-based UI is central to the overall user experience. Web performance can be defined as the speed at which a page loads and renders content to the user (Szalek and Borzemski 2019). Another definition is the ability to provide sufficient efficiency via a reasonable amount of resources (Khameesy, Magdi, and Khalifa 2017). From the perspective of the end-user, web performance is measured by the time between clicking the Web page link and the completion of total page downloading. This can be formally represented as:

$$\min_\rho \ \rho = \alpha_T - \beta_T \qquad (2)$$

Where ρ is the load time from the end user's perspective which is a time value (in seconds), α is the time at which the page content downloads, and β is the time at which the user opened the link. Thus, the goal of any quality web-based technology UI development would be to minimize ρ, thereby maximizing web performance. Web Performance is influenced by factors including client-server network connectivity, HTTP protocol version, network loss and delay, server load, and delays in domain name resolution (Iyengar et al. 2002). In numerous modern web-based applications, cloud-based technology plays a central role and as such optimal web performance is a highly desirable feature (Shivakumar 2020). Studies have revealed a strong correlation between slow web pages and revenue loss as a result of user dissatisfaction (Wijnants et al. 2018). Also, research has concluded that web performance influences customer retention and product/service acquisition (Uzayr 2022). A quick web UI load speed is synonymous to more visitors willing to return and firms such as Google and Amazon lost 20% in revenue due to half a second increase in page load time, and 1% decrease in sales due to an additional load time of 100 milliseconds respectively (Uzayr 2022). Thus web performance is relevant to site traffic and business revenue (Marx 2018).

Studies on Web Performance Optimization (WPO) have recommended the following as means of optimal performance, namely: Caching (via Browser caching, Asset caching, CDN Caching, Page caching, and custom caching) and Prefetching (achievable via Association rules-based prefetching, Markov prediction model, prediction by partial match, Clustering techniques, Stochastic Petri nets, Effective Cost Functions, dependency graph, and many more) (Shivakumar 2020). Also, A WPO checklist (Categorized Performance Rules) has been constructed and consists of Request Optimization, Web object size optimization, HTTP Header Optimization, Asset placement, Image Optimization, Network Optimization, External Dependency Optimization, and Web Application design optimization (Shailesh and Suresh 2017). A study revealed how WPO has contributed to an increase in conversion rate from 6.35% to 14.30% within a period of six months (Szalek and Borzemski 2019). In addition to that, best practices such as asynchronous alternatives, iterative testing, lightweight design, search-centered experiences, omni-channel optimization, layered architecture, and monitoring have been encouraged. Thus, indicative of the progress being made in research with regards to optimizing web performance (Shivakumar 2020).

A number of Web Performance pitfalls identified by studies include: (1) Redirects account for a significant share of Page Load Time (PLT) and a substantial amount of user-centric load time metrics such as Time-To-First-Paint (TTFP) - i.e. the time when the browser flushes the DOM's current state to the rendering engine; (2) Browser bugs influence the inaccuracy of HTTP Archive (HAR) body size for a significant number of objects; (3) A lack of data sources (a lack of DNS response which is as a result of requests not being able to establish a TCP connection, as well as the existence of certificate errors) (Enghardt, Zinner, and Feldmann 2019). As reported by research, a challenge lies in the quantification of web performance due to the diversity of web pages, the heterogeneity of devices and browsers, the choice of metrics (network-centric, browser-centric, as well as user-centric metrics), and the lack of effective standards (Enghardt, Zinner, and Feldmann 2019).

### 4.2 Phase 2 – User-Based Evaluation (Usability)

The third metric (which is the user-based evaluation), web usability, is defined as the extent to which a product or system can be utilized by specific users in order to achieve desired goals effectively, efficiently and satisfactorily within a given context of specific use. System usability is characterized by: understandability and suitable for user needs; learnability; ease of operation and control; prevent users from making mistakes; attractive UI that satisfies user interaction; allows it to be used by users with certain characteristics and disabilities. Web UI Usability evaluation is essential to system improvement. Usability evaluation is categorized into expert-based or user-based testing methods. Research has highlighted that usability cannot be directly measured (Ruscher et al. 2016), as such industry practitioners utilize numerous usability techniques such as Questionnaire for user interface satisfaction (QUIS); Software usability measurement inventory (SUMI); System usability scale (SUS); User experience questionnaire (UEQ); Computer System Usability Questionnaire (CSUQ); Usability Metric for User Experience (UMUX); NetQu@l (Mahmud et al. 2020). The survey format of user-based testing is adopted by this study.

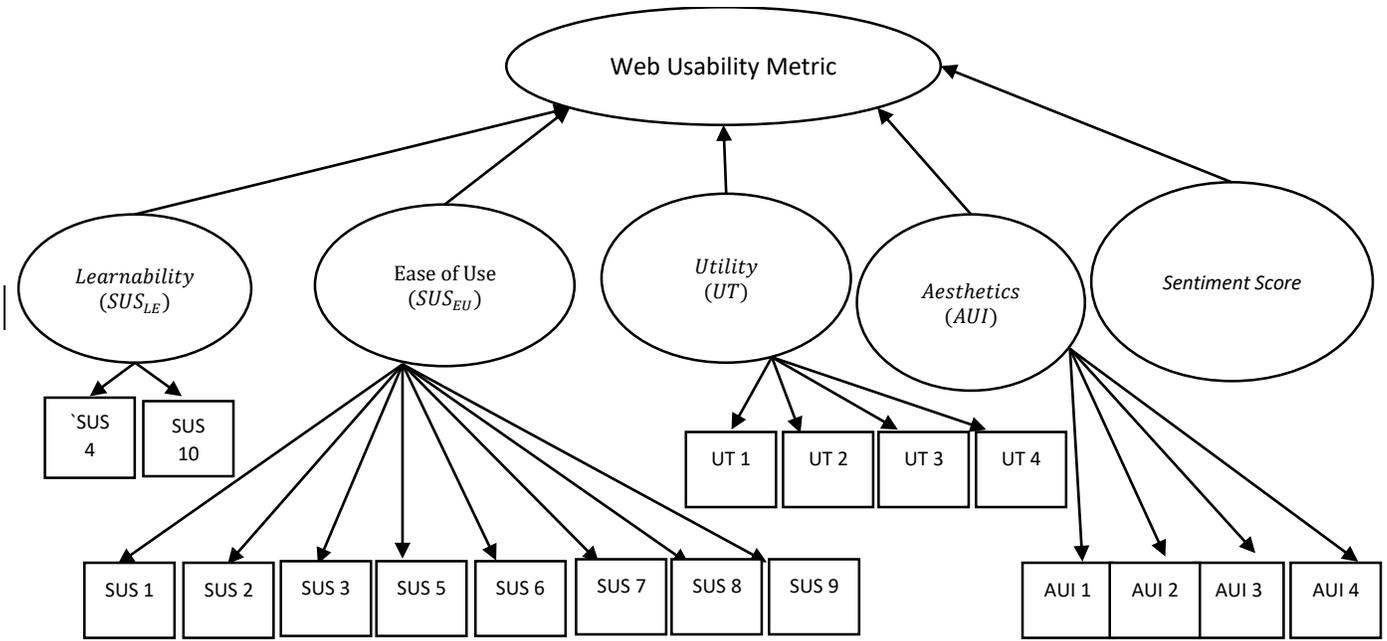

Fig. 6 Overall Usability Score – Extended System Usability Scale

Another crucial component of the usability score is user reviews. UX Studies have utilized mandatory feedback in e-learning systems and observed a distinct increase in the quality of feedback. As such, this study adopted the mandatory feedback component in UI-UX quality evaluation, since evaluations are conducted over longer timeframes (for example, after system updates, over a number of months, or yearly). Sentiment analysis is adopted as a technique to evaluate user reviews. Each review text from the user evaluation form undergoes text preprocessing and is passed through a 2-dimensional tensor (tensor2d) with padded sequences which allows for batch processing. This is followed by applying the sentiment_cnn_v1 metadata word embedding model (with max_length: 100, vocabulary_size: 20000, and embedding size: 128) and finally, the sentiment score is then calculated (from 0 to 1). The sentiment score contributes to the usability component of the overall UI Evaluation Model as represented below:

$$\sigma(z)_j = \frac{e^{z_j}}{\sum_{k=1}^{K} e^{z_k}} \; for \; j = 1, \dots, K \qquad (3)$$

Where:

δ – Softmax

$\vec{z}$ – Input vector

$e^{z_i}$ – Standard exponential function for input vector

K – Number of classes in the multi-class classifier

$e^{z_j}$ – Standard exponential function for output vector

The extended SUS survey is submitted to users of the web-based UI and the usability for user *i*, is calculated as follows:

$$ ŭ_i = \frac{\left(5\left(\sum_{o=\{1,3,5,7,9\}}(uq_o) + \sum_{i=11}^{17} uq_i - 12\right) + \left(25 - \sum_{e=\{2,4,6,8,10\}}(uq_e)\right)\right) + \left(\frac{\sigma(z)}{10}\right)}{3.5} \quad (4) $$

Where:

$uq_o$ – Odd-numbered questions

$uq_e$ – Even-numbered questions

$uq_i$ – Extended questions (system aesthetics and utility)

From the perspective of this research, Usability is measured via user-based testing. The study adopts the System Usability Scale (SUS) and extends it to encompass several shortcomings such as the lack of measures for aesthetics and system utility. The SUS was chosen as the appropriate user-based testing component of the evaluation framework because studies have supported its ability to be implemented within a timeframe of 60 seconds without imposing additional unsolicited cognitive load on end users (Mahmud et al. 2020). Extending SUS (with subcomponent learnability), the research proposed the addition of user sentiment (open-ended question), system utility, and aesthetics and UI structure (as seen in Figure 6).

### 4.3 Phase 3 – Formalization of UI-UX Evaluation (Formula and Coefficient Generation)

A key component in the formula is the formulation of the coefficients of the three (3) core metrics. As such, mathematical modelling for linear system of equations is utilized. Mathematical modelling is a description of a system using mathematical concepts and language used to explain a system and to study the effects of different components. It has been recognized as a tool to aid in quantifying and interpreting the corresponding data for a given scenario. In this study, mathematical modelling is utilized in the formulation of an evaluation metric for web UI-UX quality (WUIQ) by adopting Multi-Criteria Decision Making (MCDM) – Analytical Hierarchical Process (AHP), Statistical Analysis for testing qualitative data robustness, and K-Means Cluster Analysis for stratifying WUIQ results by user classifications. Upon identifying the three (3) metrics, a linear equation which calculates the web UI-UX quality (in percentage), demanded the use of mathematical modelling (using AHP) to generate the coefficients/weights of the individual parameters (web accessibility ($\omega_A$), web performance ($\omega_P$), and web usability ($\omega_U$)).

To undertake the process of generating coefficients/weights for the chosen web UI-UX quality metrics, AHP is adopted in this procedure. As such experts' opinions are involved in the process. MCDM has been applied in the evaluation of websites of Iranian universities (VIKOR method) - based on six (6) dimensions – "usability", "content", "functionality", "efficiency", "student services", and "reliability" – to expose design weaknesses thus improving UI and UX (Hosseinpouli Mamaghani and Montazer 2021). Another study evaluated websites of renowned higher educational institutes (fuzzy AHP methodology) to measure the usability of these academic websites (Muhammad et al. 2021). A study also used MCDM (AHP and TOPSIS methods) to identify the required factors to consider when improving customer experience in e-commerce (Davidavičienė, Markus, and Davidavičius 2020). Another study adopted the Logarithmic Fuzzy Preference Programming (LFPP) method to rank Indonesian university websites based on accessibility and usability criteria - backlink, stickiness, and web page loading time (Wardoyo and Wahyuningrum 2018). These studies give proof to the efficacy of MCDM as a tool in UI research.

MCDM algorithms are highly recommended, in spite of MCDM's known challenges such as the lack of standard decision making speed and accuracy, potential inaccuracy in results (due to its simplicity) (Dvorskỳ, Pavlenko, and Kopřiva 2015; Nasiri, Nasiri, and Azar 2016), as well as its subjectivity. This study employed the pairwise comparison method of AHP (Analytical Hierarchical Process). The reason for adopting multi-criteria decision making was due to MCDM's ability to evaluate all given options under variable degrees as well as its ease of use and capacity to handle multiple inputs and outputs (Kollati and Debnath 2021; Wang and Pang 2011). Also, no research has utilized MCDM as a medium for coefficient generation (i.e. mathematical modelling) with respect to web UI-UX quality.

AHP has been applied by researchers as a meta-heuristics strategy to solve the issue of selection of suppliers; select the most appropriate vehicle for consumers (Ghosh, Chakraborty, and Dan 2012). With respect to UI research, MCDM, particularly AHP (and Fuzzy AHP in other scenarios) has been applied in the evaluation of remote video conferencing software based on security, usability, functionality, technical performance, and pricing; usability of information services based on the following criteria attractiveness, learnability, navigability, understandability, and searchability; prioritizing usability heuristics according to the following top 10 criteria - "user control and freedom, visibility of system status, consistency and standards, error prevention, compatibility between the system and the real world, recognition rather than recall, help and documentation, flexibility and efficiency of use, aesthetic and minimalist design, helping users identify, diagnose and recover from errors" (Poositaporn, Onuean, and Jung 2022; Toan, Dang, and Hong 2022).

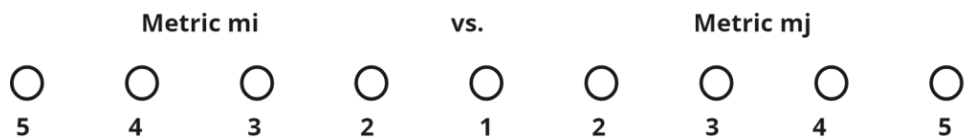

Fig. 7. Sample of Pairwise Metric Comparison in Survey

In the case of this research, an expert is defined as a stakeholder of a given web platform who is knowledgeable in the realm of web design and development, UI/UX design, as well as UI/UX quality assurance (QA). The expert survey is undertaken by expert stakeholders (UI designers and developers, backend developers, web quality assurance, and web engineers). A 5-point scale is utilized for assessing the importance of metrics over each other (Figure 7). AHP is justified as a tool for coefficient formulation based on expert opinion (pairwise comparison) due to the fact that decision-making involves a good deal of subjectivity as purported by research.

| Standard Values ($m_{ij}$) | Inverse Values ($\frac{1}{m_{ij}}$) | Definition |
|---|---|---|
| 1 | 1 | Equal Importance |
| 3 | 1/3 | Low Importance |
| 5 | 1/5 | Strong Importance |
| 2, 4 | ½, ¼ | Intermediate Values |

Table 1. Pairwise Comparison Values

As seen in Figure 5, the expert undertakes a series of pairwise comparison based on subjective judgment of the relative importance of the alternatives within the same hierarchical level. The individual pairwise comparison results are transformed into numbers (via ratio scales); which are later on placed in a new matrix for further establishing decision weights (i.e. priorities for alternatives). The Saaty Eigenvector method (based on the principal eigenvector) is used to estimate priorities within a judgment matrix.

| n | 1 | 2 | 3 | 4 | 5 | 6 | 7 | 8 | 9 | 10 |
|---|---|---|---|---|---|---|---|---|---|---|
| RI | 0.00 | 0.00 | 0.58 | 0.9 | 1.12 | 1.24 | 1.32 | 1.41 | 1.45 | 1.49 |

Table 2. Average Random Consistency Index (Saaty 1980)

After identifying the decision problem, and determining the goal, in this study, the AHP methodology is as follows:

a.  The framework consists of the goal (i.e. WUIQ), with the intermediate orders (the metrics: Performance, Accessibility, and Usability).

b.  Creation of a pairwise comparison matrix (n × n) using a Saaty's 1–5 scale of absolute numbers where the pairwise comparison matrix is defined by which item (metric) outperforms the other. Table 2 highlights the values used during the pairwise comparison.

c.  Determination of the pairwise comparison's significance by creating a relative ranks matrix.

d.  Calculation of the consistency index (CI) and consistency ratio (CR).

e.  Exclusion of weights that do not meet the CI and CR criteria.

f.  Geometric Mean of weights that meet the criteria.

For a given matrix ($M_e$, where, the pairwise comparison choices by each expert for the degree of *N=3*, where *N* is the number of metrics to be compared is represented as a follows:

$$M_e = \begin{pmatrix} m_{11} & m_{12} & \cdots & m_{1n} \\ \vdots & \vdots & \cdots & \vdots \\ m_{n1} & m_{n2} & \cdots & m_{nn} \end{pmatrix} \qquad (5)$$

$$\acute{M}_e = \begin{pmatrix} \frac{m_{11}}{\sum_{i=1}^{n} m_{11}} & \frac{m_{12}}{\sum_{i=1}^{n} m_{12}} & \cdots & \frac{m_{1n}}{\sum_{i=1}^{n} m_{1n}} \\ \vdots & \vdots & \cdots & \vdots \\ \frac{m_{n1}}{\sum_{i=1}^{n} m_{n1}} & \frac{m_{n2}}{\sum_{i=1}^{n} m_{n2}} & \cdots & \frac{m_{nn}}{\sum_{i=1}^{n} m_{nn}} \end{pmatrix} = \begin{pmatrix} \acute{m}_{11} & \acute{m}_{12} & \cdots & \acute{m}_{1n} \\ \vdots & \vdots & \cdots & \vdots \\ \acute{m}_{n1} & \acute{m}_{n2} & \cdots & \acute{m}_{nn} \end{pmatrix} \quad (6)$$

$$W_e = \begin{pmatrix} \frac{\acute{m}_{11} + \acute{m}_{12} + \cdots + \acute{m}_{1n}}{n} \\ \vdots \\ \frac{\acute{m}_{n1} + \acute{m}_{n2} + \cdots + \acute{m}_{nn}}{n} \end{pmatrix} = \begin{pmatrix} w_1 \\ \vdots \\ w_n \end{pmatrix} \quad (7)$$

(a) $M_e = N \times N$, where $e$ represents an expert's pairwise comparison

(b) There exists a diagonal with value 1; where $M_e(i,i) = 1$ for each $i, 1 \leq i \leq N$.

(c) $m_{ij} > 0$, $m_{ij} = (m_{ji})^{-1}$, $\forall$ i and j $\in$ {1, 2, 3, …, v}

(d) An entry row *i* and column *j*, is an integer $m_{ij}$, $1 \leq m_{ij} \leq 5$, or its inverse $\frac{1}{m_{ij}}$ ; Also, $m_{ij} = (m_{ji})^{-1}$ (Table 1).

The pairwise comparison matrix for a given expert ($M_e$) is furthermore normalized into ($\acute{M}_e$) by dividing each entry ($m_{ij}$) by the sum of columns. The row-wise sum of normalized pairwise comparisons is performed to derive the coefficients;

$$\text{where, } \boldsymbol{W_e} = \{\boldsymbol{\omega_P}, \boldsymbol{\omega_A}, \boldsymbol{\omega_U}\}. \quad (8)$$

$$\boldsymbol{CI} = \frac{\lambda_{max} - N}{N - 1} \quad (9)$$

The mean random inconsistency index is computed by averaging the consistency indices CI(N) of n randomly generated pairwise comparison matrix where n is large. Where $\lambda_{max}$ — eigenvalue, and N — is the number of criteria (in this study, it is represented as the three (3) metrics).

Finally to amalgamate all previous individual algorithms, the formal mathematical representation of the web UI-UX quality evaluation formula is derived as follows:

$$WUIQ_t = \sum_{i=1}^{\infty} (P_i \omega_P) + (A_i \omega_A) + (U_i \omega_U) \quad (10)$$

$$U_i = \left( \prod_{j=1}^{n} \breve{u}_j \right)_i^{\frac{1}{n}} \quad (11)$$

$$WUIQ_t = \sum_{i=1}^{\infty} (P_i \omega_P) + (A_i \omega_A) + \left( \left( \prod_{j=1}^{n} \breve{u}_j \right)_i^{\frac{1}{n}} \omega_U \right) \quad (12)$$

Where:

$WUIQ$ – Web UI-UX quality metric for the cluster i

t – Time (in months/years)

$\omega_A$ – Web accessibility coefficient

$A$ – Web accessibility score

$\omega_P$ – Web performance coefficient

$P$ – Web performance score

$\omega_U$ – Web usability coefficient

$U$ – Web usability coefficient

To validate the hybrid evaluation framework proposed, this paragraph highlights an applied use-case based on data gathered from users of a web 3.0 non-fungible token (NFT) marketplace – whose name will remain anonymous as ordered by the owners of the platform. For experimental purposes, the study focused on a small subset of the user community via random sampling through the Discord channel of the group in August 2022. Below are the results obtained for calculating $WUIQ_t$ where t = 1 (i.e. iteration 1):

a. Coefficient formulation based on expert pairwise comparison from 11 experts (4 Fullstack Developers, 4 IT Project Managers, 2 Application Testers, and 1 UI Designer). Based on calculations from formulas (5) to (9), the metric coefficients obtained are as follows:

   i. $\omega_P = 0.36$

   ii. $\omega_A = 0.27$

   iii. $\omega_U = 0.37$

The result revealed that the system provider stakeholders value web UI-UX usability more than performance and accessibility respectively for their web information system.

b. Automated Evaluation Results derived from Google Lighthouse:

   i. Web Performance Score ($P_1$) = 25%

   ii. Web Accessibility Score ($A_1$) = 97%

c. User-Based Evaluation Results for 112 simulated respondent results calculated with the extended system usability scale formula (see equation 4):

   i. Web Usability Score ($U_1$) = 61.84%

d. Finally, the resultant web UI-UX quality score (based on formula (10)) is:

$$WUIQ_1 = \sum_{i=1}(0.25 \times 0.36) + (0.97 \times 0.27) + (0.6184 \times 0.37)$$

$$WUIQ_1 \approx 0.58 = 58\%$$

Thus scoring 58% web UI-UX quality (for time t=1) is indicative of poor performance and the web information system requires improvement followed by another set of automated and usability tests – for time t=2 – but maintaining the previously obtained coefficients as a standard.

### 4.4 Usability Segmentation (K-Means Clustering)

Furthermore, to gain insight into user responses, the next step after statistical calculations of web usability focuses on user segmentation, which involves identifying users who have similar user experience quality preferences into similar groups in order to understand the common characteristics of each group and find ways to meet the needs of these users. Segmentation is often carried out according to user demographic information such as gender, education level, technical knowledge, which lacks a comprehensive account of the overall performance of users.

The unsupervised machine learning algorithm of K-Means clustering was adopted for the study. K-means clustering classify the items based on attributes/features into K number of group. This unsupervised algorithm has been applied in clustering studies as well customer segmentation and is preferred due to its ability to easily adapt to available data by modifying the feature attributes; it can work with a high dimensionality of features; it also guarantees convergence; scales to larger datasets; less computational complexity and execution time in comparison to other clustering algorithms; minimizes within-cluster variability and maximizes between-cluster variability.

K-means classifies the items (data) into groups of K numbers based on attributes/features using either the Euclidean distance and Manhattan distance metric. The dataset is represented as $X = \{x_1, x_2, \ldots, x_n\}$, in a Euclidean space (d-dimensional) $\mathbb{R}^d$. Cluster centers (c), represented as $A = \{a_1, a_2, \ldots, a_c\}$. Furthermore, let $z = [z_{ik}]_{nxc}$, where $z_{ik}$ is a binary variable ($z_{ik} \in \{0,1\}$) representative of the data point $x_i$ belongs to cluster k ($k = \{1, \ldots, c\}$). Thus, the objective function of the k-means algorithm is represented as $J(z, A) = \sum_{i=1}^{n} \sum_{k=1}^{c} z_{ik} \|x_i - a_k\|^2$. The algorithm is iterated in order to minimize the objective function $J(z, A)$ by updating the cluster centers and its member data points respectively, as:

$$a_k = \frac{\sum_{i=1}^{n} z_{ik} x_{ij}}{\sum_{i=1}^{n} z_{ik}} \qquad (13)$$

$$z_{ik} = \begin{cases} 1, & if \|x_i - a_k\|^2 = \min_{1 \leq k \leq c} \|x_i - a_k\|^2 \\ 0, & otherwise. \end{cases} \qquad (14)$$

$$SSE = \Sigma_{i=1}^{k} \Sigma_{xj \in C_i} \|x_j - \mu_i\|^2 \qquad (15)$$

To obtain the optimal number of clusters, as there is no hard-wired rule, a number of techniques exist and the study adopted the Elbow Method. This is a heuristic in mathematical optimization where the optimal number of clusters. The Scree Plot highlights the optimal clusters (k=2) – as highlighted in figure 8.

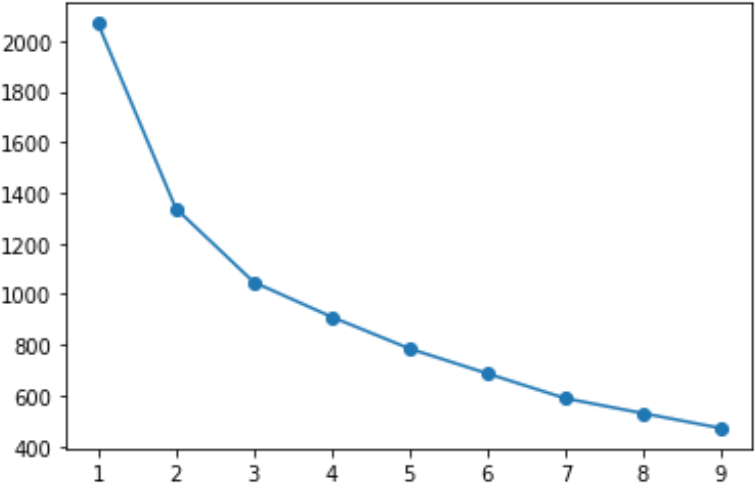

Fig. 8. Dynamic selection of the optimal number of clusters (k) using the scree plot (elbow method)

| Cluster | Duration (months) | Usability Score |
|---|---|---|
| 0 | 10.2 | 65.5 |
| 1 | 10.6 | 70.2 |

Table 3. Cluster Results – Segmented User Groups

The clusters generated from the usability segmentation via k-means is shown in Table 3. Figure 9 illustrates clusters for k=2, with usability on the y-axis and system usage duration on the x-axis.

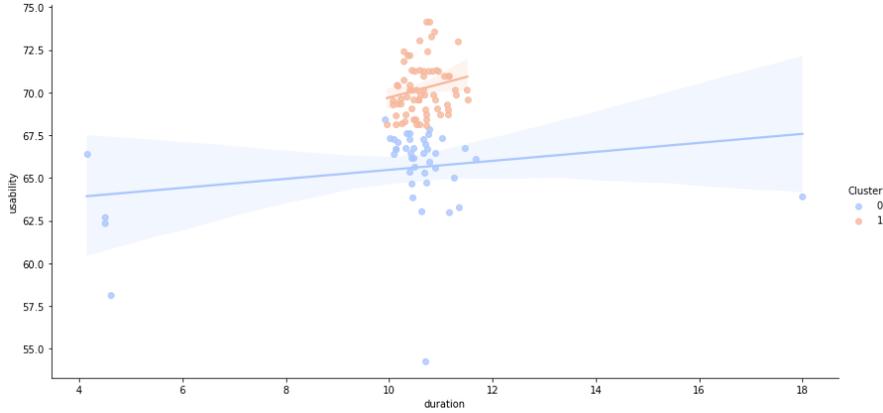

Fig. 9. Result of cluster analysis (k=2)

### 4.5 Applying Explainable AI of Web Usability Clusters

Upon obtaining the cluster model which segments users based on usability features, the final step was to interpret the cluster classification model results. Explainable/Interpretable AI and ML has recently become a key instrument in understanding the black-box ML and Deep Learning models. Explainability or Interpretability is capable of justifying AI-based decisions and it can assist in verifying predictions, for model improvement, as well as insight gaining. This leads to more trustworthy AI solutions. As such the research adopted the paradigm a tool to provide in-depth understanding to usability cluster classification per user groups.

SHAP (SHapley Additive exPlanations) is a game theoretic approach to explain the output of any machine learning model. SHAP and visualizes interpretations as SHAP summary plots and SHAP dependence plots. SHAP approximation techniques that exist include Kernel, Deep, and Tree SHAP which are used for kernel-based, deep neural network based, and tree-based models respectively.

$$\emptyset_i(f,x) = \sum_{z' \subseteq x'} \frac{|z'|!\,(M - |z'| - 1)!}{M} [f_x(z') - f_x(z'\backslash i)]$$

$$Total\ Number\ of\ Subsets = 2^n$$

(16)

The expression above derives the Shapley value ($\emptyset$) for a specific feature (*i*), for example System Utility (*s_utility*) for the blackbox model (*f*) – the segmentation model. The input datapoint which is a single sample observation (in the case of this research, a set of a user's usability choices) is represented by *x*. All possible subsets ($z'$) are iterated over, to account for interactions between individual feature values. This makes the usage of Shapley values computationally expensive for larger features. In certain scenarios $x'$ which is a simplified data input (a transformation of *x*) is utilized. A selection of subsets, for example a subset of Aesthetics (*s_aesthetics*) and System Usage Duration (*duration*) with the remaining features treated as unknown. Since the model depends on a fixed size of features for inference, the values for non-members of the subset in focus are randomly generated during the calculation of marginal contribution of a feature of interest($\emptyset_i$). The $f_x(z')$ component of the formula indicates the combination of the blackbox model and the subset with feature of interest ($\emptyset_i$). Whereas $f_x(z'\backslash i)$ represents the combination of the blackbox model and a subset without the feature of interest. The difference ($[f_x(z') - f_x(z'\backslash i)]$) is indicative of the marginal value which is defined as the contribution of $\emptyset_i$ to the subset in focus (represented in percentages). This is repeated for each permutation of subsets and weighted according to the number of players (features) in the correlation – represented by *M*. Hence the $\frac{|z'|!(M-|z'|-1)!}{M}$ component of the formula isF responsible for weighting while $[f_x(z') - f_x(z'\backslash i)]$ is responsible for measuring contribution. For experimental purposes and to understand the contribution of features to the model behavior per each cluster. Also, the feature importance based on mean absolute value of the SHAP values in the subsequent plots.

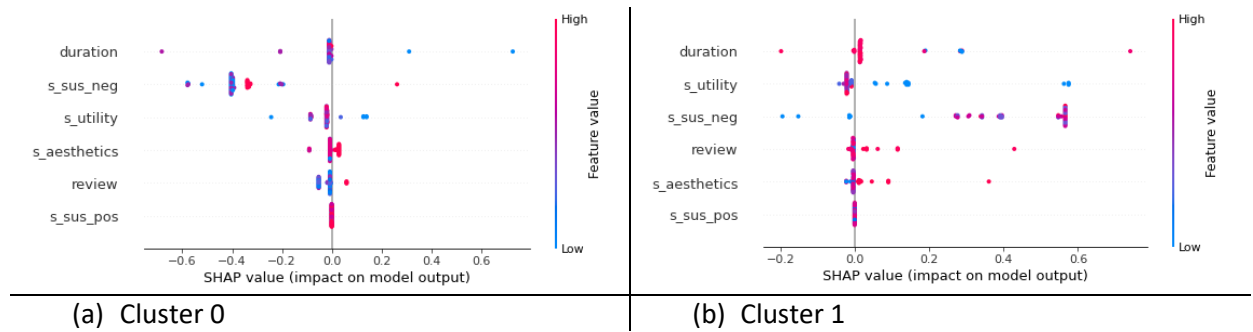

(a) Cluster 0  (b) Cluster 1

Fig. 10 (a, b) Beeswarm plot and histograms highlight the effect of features on classification.

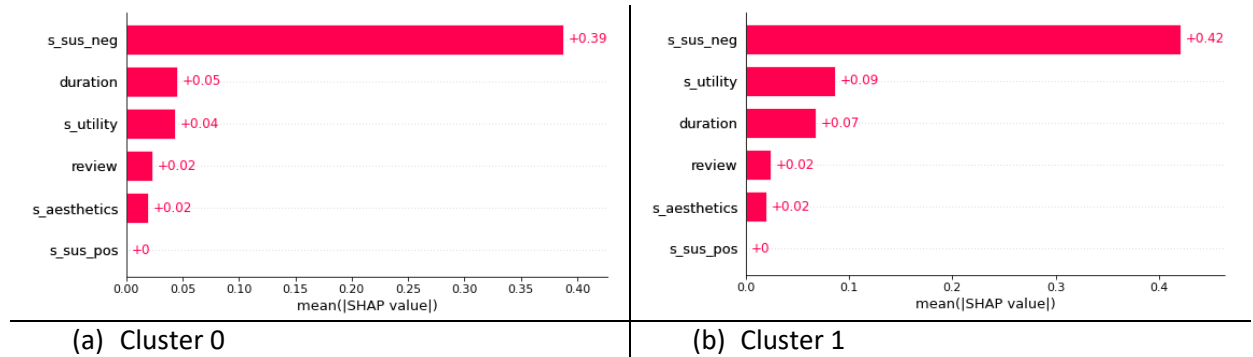

(a) Cluster 0  (b) Cluster 1

Fig. 11 (a, b). Feature Clustering (Independent Variables) Importance Plot

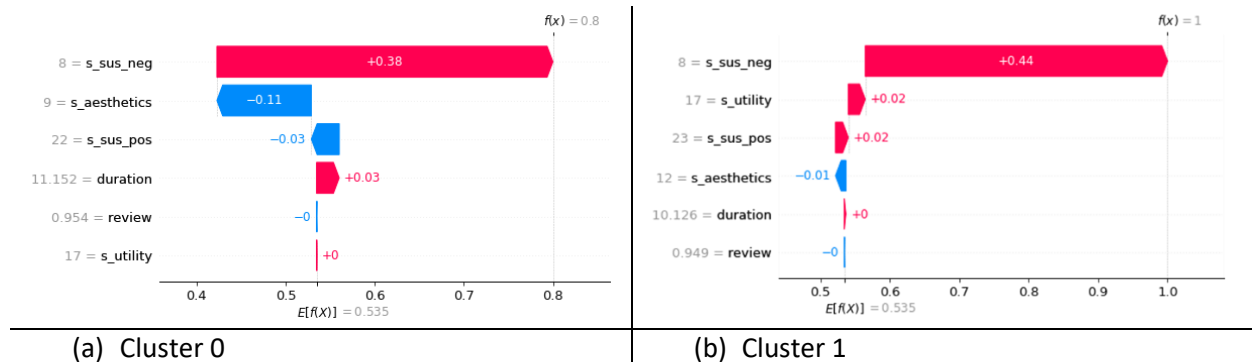

(a) Cluster 0  (b) Cluster 1

Fig. 12 (a, b) Feature Contribution to the Clusters

The image above (Рисунки 10, 11 and 12) shows the features, each of which contributes to moving the model output from the base value (the average of the model output over the training dataset) to the model output. Features that increase the forecast are shown in red, and those that reduce the forecast are shown in blue. In the case of the experiment of the data collected for the case study, where the expected value $E[f(X)] = 0.535$, the negative components of the system usability scale survey contributed primarily to both clusters, with the duration of system usage not being a contributing factor for cluster 1 while the opposite is for cluster 0. These results provide an overview for decision makers and serve as a feedback mechanism for improving upon systems and furthermore developing personalized UIs in the future.

## 5. Conclusion

The article aimed at introducing a novel hybrid multi-stakeholder centric framework (which combines a series of data-driven algorithms) to evaluate the web UI-UX of any given information system as means of supporting web information system design and improvement decision making. In summary, the study focused on three key performance indicators for assessing the web UI-UX quality namely – (1) performance and (2) accessibility – which are automatically tested, and (3) usability – which is derived via user-based testing (by extending the system usability scale questionnaire to make up for system aesthetics and utility). To obtain a standardized score for web UI-UX quality, a simple linear formula is proposed with unknown coefficients for each of the three (3) KPIs. The unknown KPI coefficients are formulated by using the analytical hierarchical process (AHP) methodology based on the geometric mean of pairwise comparisons from experts (web designers, developers, quality assurance engineers, and user experience specialists). Upon obtaining the coefficients, the KPI values are multiplied with their respective coefficients to obtain the web UI-UX quality score. Furthermore, to comprehend the user's perspectives towards the given system, usability segmentation is then conducted using k-means clustering algorithm followed by the application of explainable artificial intelligence to deeply understand the various usability clusters. It is believed that these in-depth techniques will provide designers and developers with user-centric perspectives for improving the UI and UX quality of web-based information systems.

The theoretical contribution of this study is evident in the inclusion of mathematical modelling, multi-criteria decision making and hybridization of user-based and automated web testing techniques in evaluating the quality of UI and UX of web-based information systems. From a practical standpoint, the study contributes to the monitoring phase of the DevOps process of web development by proposing the hybrid evaluation framework. Just like every other research has limitations, a limitation of the proposed framework is its lack of direct backend-related evaluation such is business logic and algorithm speed, as well as not paying attention to latent components that may affect web UI-UX.

Future research will seek to deploy a one-stop shop for realizing the framework and managing the UI-UX quality journey of web information systems. It is recommended for future researchers to also identify the feasibility of integrating state-of-the-art artificial intelligence techniques in the framework to optimize the assessment procedure.

**Conflict of Interest**